\documentclass[
 reprint,
 amsmath,amssymb,superscriptaddress,
 aps,pra
]{revtex4-1}

\usepackage{graphicx}
\usepackage{dcolumn}
\usepackage{bm}
\usepackage{hyperref}

\hypersetup{colorlinks=true, citecolor=blue}
\usepackage[caption=false]{subfig}
\usepackage{physics}
\usepackage{bbm}
\usepackage{amsmath}

\usepackage{ulem}

\begin{document}


\title{Many-body probes for quantum features of spacetime}
\author{Hadrien Chevalier}\thanks{hadrien.chevalier17@imperial.ac.uk}
\affiliation{QOLS, Blackett Laboratory, Imperial College London, SW7 2AZ, United Kingdom}
\author{Hyukjoon Kwon}
\affiliation{QOLS, Blackett Laboratory, Imperial College London, SW7 2AZ, United Kingdom}
\affiliation{Korea Institute for Advanced Study, Seoul, 02455, Korea}
\author{Kiran E. Khosla}
\affiliation{QOLS, Blackett Laboratory, Imperial College London, SW7 2AZ, United Kingdom}
\author{Igor Pikovski}
\affiliation{Department of Physics, Stockholm University, AlbaNova University Center, SE-106 91 Stockholm, Sweden}
\affiliation{Department of Physics, Stevens Institute of Technology, Castle Point on the Hudson, Hoboken, NJ 07030, USA}
\author{M. S. Kim}\thanks{m.kim@imperial.ac.uk}
\affiliation{QOLS, Blackett Laboratory, Imperial College London, SW7 2AZ, United Kingdom}
\affiliation{Korea Institute for Advanced Study, Seoul, 02455, Korea}
\date{\today}

\begin{abstract}
Many theories of quantum gravity can be understood as imposing a minimum length scale the signatures of which can potentially be seen in precise table top experiments. In this work we inspect the capacity for correlated many body systems to probe non-classicalities of spacetime through modifications of the commutation relations. We find an analytic derivation of the dynamics for a single mode light field interacting with a single mechanical oscillator and with coupled oscillators to first order corrections to the commutation relations. Our solution is valid for any coupling function as we work out the full Magnus expansion. We numerically show that it is possible to have superquadratic scaling of a non-classical phase term, arising from the modification to the commutation relations, with coupled mechanical oscillators.
\end{abstract}

\pacs{Valid PACS appear here}
\maketitle


\section{\label{sec:intro}Introduction}
Establishing a quantum theory of gravity is one of the greatest challenges in contemporary physics. 
It is widely thought, in many different approaches to quantum gravity, that there should be a minimum length scale, typically taken to be on the order of the Planck length $l_p = \sqrt{\hbar G/c^3}$, where $\hbar$ is the reduced Planck constant, $G$ is the gravitational constant and $c$ is the speed of light in vacuum~\cite{garay1995quantum,hossenfelder2013minimal}. Because the standard Heisenberg uncertainty relations do not prevent arbitrarily fine squeezing of a state in position space, at the expense of high uncertainty in momentum space, the existence of a minimum length scale comes with a generalized uncertainty principle (GUP) $\Delta x\Delta p \geq \frac{\hbar}{2}(1 + \beta \Delta p^2)$, where $\beta = \beta_0/(M_pc)^2$ is what we shall call the GUP correction, and $M_p$ is the Planck mass~\cite{maggiore1993algebraic, ali2009discreteness}. In turn this can be viewed, using Robertson's inequalities~\cite{robertson1929uncertainty}, as resulting from a typical deformation of canonical commutation relations $[x_i,p_j] = i\hbar(1 + \beta p_j^2)\delta_{ij}$~\cite{kempf1995hilbert}. However there has yet to be empirical evidence suggesting corrections from classical commutation relations, as current observations do not rule out values of $\beta_0$ much below $10^{6}$~\cite{bawaj2015probing, bushev2019testing}, while it is assumed $\beta_0$ should be on the order of unit for Planck scale corrections~\cite{das2008universality}. Although non-classical features of spacetime and gravitation may be censored by some Penrose type objective collapse process due to gravity itself~\cite{penrose1996gravity,penrose2014gravitization, diosi1987universal}, in recent years much attention has been given to proposals seeking to uncover some non-classicalities of spacetime while circumventing the requirement to work at the Planck energy level~\cite{bose2017spin,marletto2017gravitationally,bekenstein2012tabletop,pikovski2012probing,marin2013gravitational}.
By relinquishing high-energy probes for directly observable Planckian effects, those proposals rely instead on finely controlled low-energy experiments which would allow to indirectly detect signatures of deviations from classical theory.

Here we study a toy model inspired from a proposal that we term ``optomechanical protocol''~\cite{pikovski2012probing} as well as a reaction to all protocols involving massive systems to test deformed commutators~\cite{amelino2013challenge}. The optomechanical protocol suggests that one can gain information on the type of evolution undergone by a mechanical oscillator by inspecting the non-linear phase that the light picks up after having interacted with the oscillator. A sequence of light pulses drives the mechanical oscillator around a close phase space trajectory, returning the oscillator to its initial state (in the absence of commutator deformations), while imparting a (mechanical state-independent) non-linear optical phase. This phase is of interest as it may carry non-classical signatures and, as we shall show from first principles, is quartic in the photon number if GUP corrections exist, and is quadratic otherwise.

A major theoretical challenge to these types of proposals, regardless of experimental difficulties, is that one may in principle not be able to use the center of mass coordinates of any macroscopic systems, which are collections of elementary particles, to probe deformations of the commutation relations. Following supplementary information of Ref.~\cite{pikovski2012probing} and~\cite{amelino2013challenge} if one considers the center of mass coordinates of a system of $N$ equally massive subsystems $X = \frac{1}{N}\sum_{k=1}^N x_k$ and $P = \sum_{k=1}^N p_k$ with a deformed commutator $[x_i,p_j] = i\hbar(1 + \beta p_j^2)\delta_{ij}$ then \begin{equation}\label{eq:correction}[X,P] = i\hbar\left(1 + \frac{\beta}{N^2}P^2 + \frac{\beta}{N}\sum_{k=1}^N\left(p_k^2 - \frac{P^2}{N^2}\right)\right).\end{equation}
In the quasi-rigid limit, the variance of the momentum distribution over the subsystems vanishes, so that $[X,P] \approx i\hbar\left(1 + \frac{\beta}{N^2}P^2\right)$, thus deformations to the center of mass commutator are suppressed by the (potentially macroscopic) number of constituent particles. This leads to an important question: are many-body macroscopic systems relevant probes for Planck scale behaviour?

Although the scaling of the suppression is unknown in general~\cite{kumar2020quantum}, from Eq.~\eqref{eq:correction} one can expect a scaling between $1/N$ (for uncorrelated systems) and $1/N^2$ (for rigid systems). However, in specially correlated fine-tuned systems, it may be that some Planck scale corrections could become amplified: this would be the case if some non-classical signature depending on the GUP correction is found to have a superquadratic scaling with the number of elementary constituents. In our work, this non-classical signature arising from the interaction between a single mode optical field with a many-body system will be taken to be a phase term that is quartic in the photon number and is independent of the state of the many-body mechanical system. In order to observe superquadratic scaling, one must first be able to fully characterize the dynamics of many-body systems in a sufficiently general manner to allow for a broad type of interaction patterns and couplings between subsystems. This accurate description is what we undertake in this work, and what we use to show numerical evidence of superquadratic scaling.

This paper is structured as follows. In Sec.~\ref{sec:single} we present the basic situation on which we build the rest of our work, we go through classical calculations in the case of a single particle interacting with a single mode optical field and show how the dynamics is solved with an order $2$ Magnus expansion. In Sec.~\ref{sec:GUPsingle} we modify the commutation relations and completely solve the full time-dependent dynamics for arbitrary interaction functions, to first order in the GUP corrections. We observe that the order $5$ Magnus expansion gives the exact propagator, and includes a phase term that is independent of the state of the mechanical oscillator, with quartic dependence on the photon number. In Sec.~\ref{sec:Pikowski} we impose a specific form for the interaction function and  show how to recover the phase term from first principles.  We extend our analysis in Sec.~\ref{sec:many} to many-body systems and lay out a general derivation of the fifth Magnus term. In Sec.~\ref{sec:Pikovskiextended} we apply our general formula to a many-body system interacting with pulsed light, and demonstrate a numerical example of superquadratic scaling behaviour when the system is a chain of coupled harmonic oscillators.

\section{\label{sec:single}Single particle toy model}
Let us consider a single particle of mass $m$ that is harmonically trapped with angular frequency $\Omega$, interacting with a single mode optical field with annihilation operator $a$ and angular frequency $\Omega_F$. In standard quantum mechanics, the optomechanical Hamiltonian reads
\begin{equation}
    H^s = \frac{{p}^2}{2m} + \frac{1}{2}m\Omega^2x^2 + \hbar\Omega_F\left(a^{\dagger}a + \frac{1}{2}\right) + g(t)a^{\dagger}ax,
\end{equation}
where $x$ is the position of the particle with respect to its rest position, ${p}$ is its conjugate momentum, the $s$ superscript stands for ``standard", and where $g(t)$ is the coupling function. Note that in our approach we keep $g(t)$ general. By rescaling the variables as $q = \sqrt{{m}{\Omega}}x$ and $p \to \frac{1}{\sqrt{m\Omega}}{p}$ we rewrite the previous equation as
\begin{equation}
\begin{aligned}
    H^s 
    &= \frac{\Omega}{2}\left(p^2 + q^2\right) + \hbar\Omega_F\left(a^{\dagger}a + \frac{1}{2}\right) + \frac{1}{\sqrt{m\Omega}}g(t)a^{\dagger}aq.
\end{aligned}
\end{equation}
We shall define $H_1^s(t) = \frac{1}{\sqrt{m\Omega}}g(t)a^{\dagger}aq$ as the interaction part, and the free part as 
$H_0 = \frac{\Omega}{2}\left(p^2 + q^2\right)+ \hbar\Omega_F\left(a^{\dagger}a+ \frac{1}{2}\right)$.

We seek to find the resulting oscillator-state-independent phase accumulated by the light field at the end of the interaction. The fact that this phase term does not depend on the oscillator state is important as we seek to test modifications to the commutation relations. With no knowledge of the true commutation relations, the initial (ground) state of the mechanical oscillator is ill-defined. In the interaction (or Dirac) picture, operators $A$ are transformed as $\tilde A(t) = e^{iH_0 t/\hbar} A e^{-iH_0 t/\hbar}$. The unitary evolution of the system can be written as $\tilde{\rho}(t) = \tilde{U}(t)\tilde{\rho}(0)\tilde{U}^{\dagger}(t)$, and since its time derivative must be consistent with the Liouville equation $\frac{d}{dt}\tilde{\rho}(t) = -\frac{i}{\hbar}[\tilde{H}_1(t),\tilde{\rho}(t)]$, this requires the unitary propagator to satisfy $\frac{d}{dt}\tilde{U}(t) = \frac{-i}{\hbar}\tilde{H}_1(t)\tilde{U}(t)$. Since the interaction Hamiltonians do not necessarily commute at different times, the propagator solving the equation takes the form $\tilde U_t = \exp(\Theta(t))$ where $\Theta(t) = \sum_{k=1}^{+\infty} \Theta_k(t)$ is given by a Magnus series~\cite{magnus1954exponential}.

Because the first order unequal time commutators are scalars, only the two first terms in the Magnus expansion are non-vanishing. Explicitly we have
\begin{equation}
    \begin{aligned}
        \Theta_1(t) &= -\frac{i}{\hbar}\int_0^t dt_1 \tilde{H}_1(t_1) \\
        \Theta_2(t) &= \frac{1}{2}\left(\frac{-i}{\hbar}\right)^2\int_0^tdt_1\int_0^{t_1}dt_2[\tilde{H}_1(t_1),\tilde{H}_1(t_2)].
    \end{aligned}
\end{equation}
To lighten the notations, let us rewrite the interaction term as
$\tilde{H}_1^s(t) = a^{\dagger}a(g'(t)q + g''(t)p)/\sqrt{m\Omega}$,
where
$g'(t) = g(t)\cos(\Omega t)$ and $g''(t) = g(t)\sin(\Omega t).$
The first term in the Magnus expansion then reads
\begin{equation}
    \Theta_1(t) = -\frac{i}{\hbar}\frac{1}{\sqrt{m\Omega}}{a}^{\dagger}{a} (G'(t)q + G''(t)p),
\end{equation}
where $G' = \int_0^t g'(t_1)dt_1$, $G'' = \int_0^t g''(t_1)dt_1$, and the second term reads
\begin{equation}
\begin{aligned}
    {\Theta}_2(t) = -\frac{i({a}^{\dagger}{a})^2}{2\hbar m\Omega}\int_0^t dt_1 (g'(t_1)G''(t_1) - g''(t_1)G'(t_1)),
\end{aligned}
\end{equation}
where we have applied the standard commutation relation $[q,p] = i\hbar$.
Hence we find that the propagator takes the form
\begin{multline}\label{eq:ClassicalPropagator}
   \tilde{U}(t) = \exp(-\frac{i}{\hbar} \frac{({a}^{\dagger}{a})^2}{m\Omega}F(t))\\\times\exp(-\frac{i}{\hbar}\frac{{a}^{\dagger}{a}}{\sqrt{m\Omega}} (G'(t)q + G''(t)p)),
\end{multline}
where ${F}(t) = \frac{1}{2}\displaystyle\int_0^t dt_1 (g'(t_1)G''(t_1) - g''(t_1)G'(t_1))$ gives a phase term which is independent of the mechanical oscillator's state and is quadratic in the photon number.

\section{The GUP modification for a single particle}\label{sec:GUPsingle}

We now introduce the GUP modification to the commutation relations. While it is possible to analytically work out the dynamics of the system by imposing $[x,p] = i\hbar(1 + \beta p^2)$, we instead work with the usual commutation relations but redefine the momentum as ${p} \to {p}(1 + \frac{\beta}{3}{p}^2)$, as in Refs.~\cite{brau1999minimal,das2008universality}, which yields the modified commutation relations. The Hamiltonian, with reduced variables, now takes the form
\begin{multline}
    H = \frac{\Omega}{2}\left(p^2 + q^2\right) + \hbar\Omega_F\left(a^{\dagger}a + \frac{1}{2}\right) \\ + \frac{1}{\sqrt{m\Omega}}g(t)a^{\dagger}aq + \frac{1}{3} m\Omega^2 \beta p^4 + O(\beta^2).
\end{multline}
We shall now define $H_1(t) = \frac{1}{\sqrt{m\Omega}}g(t)a^{\dagger}aq + \frac{1}{3}m\Omega^2 \beta p^4$ as the interaction term. In the same manner as the standard phase was found, we work in the interaction picture where the propagator satisfies the differential equation $\frac{d}{dt}\tilde U(t) = -\frac{i}{\hbar}\tilde{H}_1(t)\tilde U(t)$. In order to compute the Magnus terms, one needs to express the different commutators between $\tilde{H}_1$ at different times, and while the $p^4$ term does give rise to many non-vanishing higher-order commutators, to first order in the GUP correction $\beta$, the Magnus expansion only has $5$ non-vanishing terms. We have explicitly
\begin{equation}\label{eq:InteractionHamiltonian}
    \tilde{H}_1(t) = \frac{1}{\sqrt{m\Omega}}g(t)a^{\dagger}a\tilde{q}(t) + \frac{1}{3}m\Omega^2\beta\tilde{p}^4(t),
\end{equation}
where usual harmonic oscillator dynamics give $\tilde{q}(t) = \cos(\Omega t)q + \sin(\Omega t)p$ and $\tilde{p}(t) = \cos(\Omega t)p - \sin(\Omega t)q$. Using standard commutation relations, one establishes the following commutator expressions,
\begin{equation}\label{eq:simplecommutators}
    \begin{aligned}
     {}   [\tilde{q}(t_1),\tilde{q}(t_2)] &= i\hbar\sin(\Omega(t_2-t_1)),\\
        \forall n\in\mathbb{N}^*, \ [\tilde{q}(t_1),\tilde{p}^n(t_2)] &= ni\hbar\tilde{p}^{n-1}(t_2)\cos(\Omega(t_2-t_1)).
    \end{aligned}
\end{equation}
To order $1$ in $\beta$, it is clear that the first commutator $[\tilde{H}_1(t_1),\tilde{H}_1(t_2)]$ will take the form $C + \beta \tilde{p}^3 + O(\beta^2)$, where $C$ commutes with the interaction Hamiltonian. So the next order nested commutators are all at least order $1$ in $\beta$. After some calculations, see appendix~\ref{sec:nestedcommutators} for details, one arrives at the following form of the fifth order commutator:
\begin{multline}\label{eq:FourNestedCommutator}
     [\tilde{H}_1(t_1),[\tilde{H}_1(t_2),[\tilde{H}_1(t_3), [\tilde{H}_1(t_4), \tilde{H}_1(t_5)]]]] = \\ \frac{4!}{3}(a^{\dagger}a)^4(i\hbar)^4 \beta \frac{1}{m} \prod_{j=1}^4 g(t_j)\cos(\Omega(t_5-t_j))- (t_4 \leftrightarrow t_5),
\end{multline}
where $(t_4\leftrightarrow t_5)$ signifies that there is an extra term that is identical up to a swap between two time arguments.
  
The fifth order Magnus term, which is an integral of such fifth order nested commutators (up to some permutations of time indices), is particularly interesting as it is independent of the mechanical-oscillator operators $\tilde{q}(t)$ and $\tilde{p}(t)$. In order to further work out its form, we note that it is a quintuple nested integral of $22$ nested commutators~\cite{prato1997note} as shown explicitly in Appendix~\ref{sec:PratoLamberti}. Simplifications are possible as $18$ of those commutators are of the form $[[\tilde{H}_1(t_i), \tilde{H}_1(t_j)], [\tilde{H}_1(t_k), [\tilde{H}_1(t_l),\tilde{H}_1(t_m)]]]$, and looking at Eqs.~\eqref{eq:CommuteOnce} and~\eqref{eq:CommuteTwice}, we see that they are at least quadratic in $\beta$. To first order in $\beta$ we are left with the four terms
\begin{multline}\label{eq:MagnusTermGeneral}
    \Theta_5(t) = \left(\frac{-i}{\hbar}\right)^5\sum_{\sigma\in S} \int_{(5,t)}dt^5 \lambda_\sigma [\tilde{H}_1(t_{\sigma(1)}), \\ [\tilde{H}_1(t_{\sigma(2)}), [\tilde{H}_1(t_{\sigma(3)}), [\tilde{H}_1(t_{\sigma(4)}), \tilde{H}_1(t_{\sigma(5)})]]]],
\end{multline}
where we have defined the nested integral operator $\int_{(5,t)}dt^5 := \int_0^t dt_1\int_0^{t_1}dt_2\int_0^{t_2}dt_3\int_0^{t_3}dt_4\int_0^{t_4}dt_5$, the set $S$ which is made up of the four permutations $(\sigma_1, \sigma_2,\sigma_3, \sigma_4) =((54321), (15423), (14325), (15324))$ and the associated coefficients $\lambda_{\sigma_1} = \lambda_{\sigma_3} = \lambda_{\sigma_4} = -\frac{1}{30}$ and $\lambda_{\sigma_2} = \frac{2}{15}$. Combining the previous equation with Eq.~\eqref{eq:FourNestedCommutator} one arrives at the following expression which holds for arbitrary coupling functions $g(t)$, to first order in $\beta$:
\begin{multline}\label{eq:GeneralMagnusFifth}
    \Theta_5(t) = -\frac{i}{\hbar}\frac{4!}{3}(a^{\dagger}a)^4\frac{\beta}{m}\sum_{\sigma\in S}\int_{(5,t)}dt^5 \lambda_{\sigma}\\ \left(\prod_{j=1}^4 g(t_{\sigma(j)})\cos(\Omega(t_{\sigma(5)} - t_{\sigma(j)})) - (t_{\sigma(4)} \leftrightarrow t_{\sigma(5)}) \right).
\end{multline}
Similar to the pulsed regime considered in Ref.~\cite{pikovski2012probing}, this fifth Magnus term, which is proportionnal to the GUP correction $\beta$, contributes to the propagator as a highly non-linear phase factor depending on the fourth power of the photon number. This is to compare with the standard case of unmodified commutation relations, where the analogous oscillator-state-independent phase factor $\exp(-\frac{i}{\hbar}m({a}^{\dagger}{a})^2F(t))$ in Eq~\eqref{eq:ClassicalPropagator}, is quadratic in the photon number. Our formula is valid for any optomechanical coupling function $g(t)$. In the next section we shall analytically apply it to the case where the coupling function consists of four pulses, as was considered in~\cite{pikovski2012probing}.

\section{Phase term in the pulsed regime}\label{sec:Pikowski}
We now make use of our general formula for a single particle, with a specific coupling function which was considered in~\cite{pikovski2012probing}. To do so, it is useful to first define $f(t_1,...,t_5) = \prod_{j=1}^4\cos(\Omega(t_5-t_j))$. In the context of pulsed interaction, the coupling term is well approximated by a a sum of Dirac distributions $g(t) = \lambda\sum_{i=0}^3 \delta(t - \tau_i)$. The Dirac delta interaction leaves the mechanical oscillator's state fixed over the duration of the pulse which is identical to the approximation used for pulsed optomechanics (up to a global phase on the light). Here a succession of four Dirac pulses at instants $(\tau_i)_{0\leq i\leq 3}$ is considered, where $\lambda \geq 0$ is a coupling strength.

After some calculations shown in appendix~\ref{sec:DetailsPikovski} we find
\begin{equation}\label{eq:MagnusTermDiracs}
\begin{aligned}
&\Theta_5(t) = \frac{-i}{\hbar} \frac{4!}{3}(a^{\dagger}a)^4\frac{\beta}{m}\frac{\lambda^4}{30}\sum_{i_1i_2i_3i_4 = 0}^3\\
&\Bigg\{H(t, \tau_{i_4},\tau_{i_3},\tau_{i_2},\tau_{i_1})\left(\int_{\tau_{i_3}}^{\tau_{i_4}}ds - \int_{\tau_{i_4}}^t ds\right) \\
 &+ H(t, \tau_{i_1},\tau_{i_4},\tau_{i_3},\tau_{i_2})\left( 4\int_{\tau_{i_3}}^{\tau_{i_4}}ds - 4\int_{\tau_{i_4}}^{\tau_{i_1}}ds - \int_{0}^{\tau_{i_3}}ds\right) \\
 &+ H(t, \tau_{i_1},\tau_{i_3},\tau_{i_2},\tau_{i_4})\left( \int_{\tau_{i_3}}^{\tau_{i_1}}ds\right) \\
 &+ H(t, \tau_{i_1},\tau_{i_3},\tau_{i_4},\tau_{i_2})\left( \int_{\tau_{i_3}}^{\tau_{i_1}}ds\right) \Bigg\}f(\tau_{i_1},\tau_{i_2},\tau_{i_3},\tau_{i_4}, s),
\end{aligned}
\end{equation}
where $H(a,b,c,d,e)$ is the succession of Heaviside functions $H(a-b)H(b-c)H(c-d)H(d-e)$. 

In the particular case where the interaction instants are $\tau_k = t_0 +  k\theta =t_0 + k\frac{\pi}{2\Omega}$, where $\theta$ is a quarter of the oscillator period, and for times $t > t_0 + 3\theta$ after which all four matter-field interactions have taken place, one finds the closed form result
\begin{equation}
    \Theta_5(t) = \left(\frac{-i}{\hbar}\right) \frac{4!}{3}(a^{\dagger}a)^4\frac{\beta}{m}\frac{\lambda^4}{30\Omega}\frac{5(9\pi - 16)}{32}.
\end{equation}
We observe that this fifth-Magnus term is a constant of time for $t > t_0 + 3\theta$, which is expected as the full system undergoes free evolution. Similar to what was found in Ref.~\cite{pikovski2012probing}, this mechanical-state-independent phase factor increases as the fourth power of the optical intensity $(a^{\dagger}a)$ and the interaction strength $\lambda$. In the following sections, we will be studying how this non-classical phase factor scales with the number of mechanical oscillators.

\section{Generalization to the many-body case}\label{sec:many}
In view of the observations made in supplementary information of Ref.~\cite{pikovski2012probing} and Ref.~\cite{amelino2013challenge}, the GUP correction $\beta$ is expected to be linearly or quadratically suppressed by the size of the considered systems when they are uncorrelated or rigid. However it may well be that in some intermediate regime, where a many-body system is neither rigid nor a collection of uncoupled particles, a non-classical signature could scale superquadratically and thus overcome the suppression of $\beta$. We have taken our non-classical signature to be the fifth order Magnus term to first order in $\beta$, which is an oscillator-independent phase factor. Here we investigate the form of this non-classical phase factor for many-body systems.

Let us extend our analysis to the case of $N$ harmonically coupled identical trapped particles. The total Hamiltonian to first order in $\beta$ can be written as $H(t) = H_0 + H_1(t)$ where the free Hamiltonian is
\begin{eqnarray}
    H_0 &=& \hbar\Omega_F\left(a^{\dagger}a+\frac{1}{2}\right) +\sum_{i=1}^N\left( \frac{1}{2m}p_i^2 + \frac{1}{2}m\Omega^2x_i^2 \right)\nonumber \\
    & &~~ +  \frac{1}{2}m\Omega_c^2\sum_{i=1}^{N-1}(x_{i+1}-x_i)^2,
\end{eqnarray}
however our approach does not require a specific potential term, and holds for general potentials of the form $\frac{m}{2}x^Thx$ where $h$ is a symmetric matrix (Hessian of the potential). The interaction Hamiltonian is
\begin{equation}\label{eq:InteractionHamiltonianMultiple}
    H_1(t) = \sum_{i=1}^N g_i(t)x_i(a^{\dagger}a) + \left(\frac{\beta}{3m}\right)\sum_{i=1}^Np_i^4,
\end{equation}
where $\Omega_c$ is the coupling frequency between neighbouring oscillators.  In the same way as before, the time propagator in the interaction picture satisfies the differential equation $\frac{d}{dt}\tilde U(t) = -\frac{i}{\hbar}\tilde{H}_1(t)\tilde U(t)$. Thus the propagator is expressed as a Magnus series $\tilde U(t) =\exp(\sum_{k=1}^{+\infty}\Theta_k(t))$. We are lead to expressing nested self-commutators of the interaction Hamiltonian. Expressing the interaction Hamiltonian~\eqref{eq:InteractionHamiltonianMultiple} in the interaction picture, one arrives at an expression similar to Eq.~\eqref{eq:MagnusTermGeneral} for the fifth order Magnus term:
\begin{equation}
    \Theta_5(t) = \left(\frac{-i}{\hbar}\right)^5\int_{(5,t)}\sum_{\sigma\in S}\lambda_{\sigma}F_{\sigma},
\end{equation}
where to first order in $\beta$ one has
\begin{equation}
\begin{aligned}
    &F_\sigma = \left( \frac{\beta}{3m} \right)  (a^\dagger a)^4 \times \\ &\sum_{i_1,i_2,i_3,i_4,i_5=1}^N g_{i_1}(t_{\sigma(1)}) g_{i_2}(t_{\sigma(2)}) g_{i_3}(t_{\sigma(3)}) g_{i_4}(t_{\sigma(4)})\\
    &[\tilde x_{i_1}(t_{\sigma(1)}), [\tilde x_{i_2}(t_{\sigma(2)}), [\tilde x_{i_3}(t_{\sigma(3)}), [\tilde x_{i_4}(t_{\sigma(4)}), \tilde p_{i_5}^4(t_{\sigma(5)})]]]] \\
    &\qquad - (t_{\sigma(4)} \leftrightarrow t_{\sigma(5)}~\&~i_4 \leftrightarrow i_5).
\end{aligned}
\label{eq:F_sigma_N}
\end{equation}

In order to express the nested commutator in general, we claim that $[\tilde{x}_i(t_1),\tilde{p}_j(t_2)] := C_{ij}(t_1-t_2)$ is a $c$-number. Indeed, as shown in details in Appendix~\ref{sec:scalarcommutator}, $C_{ij}(t_1-t_2) = i\hbar\sum_{k=1}^NO_{ik}O'_{jk}\cos(\omega_k(t_2-t_1))$ where $O$ and  $O'$ are some scaled normal mode transformation matrices defined by $O = \frac{1}{\sqrt{m}}P\omega^{-1/2}$ and $O' = \sqrt{m}P\omega^{1/2}$, where the orthogonal diagonalizing transformation $P$ of $h$ satisfies $\text{diag}((\omega_i^2)_{1\leq i \leq N}) = P^ThP$. Note that $m$ is a scalar and $\omega = \text{diag}((\omega_i^2)_{1\leq i \leq N})$ is a diagonal matrix.

One can now express the nested commutators as
\begin{multline}
    [\tilde x_{i_1}(t_{\sigma(1)}), [\tilde x_{i_2}(t_{\sigma(2)}), [\tilde x_{i_3}(t_{\sigma(3)}), [\tilde x_{i_4}(t_{\sigma(4)}), \tilde p_{i_5}^4(t_{\sigma(5)})]]]]\\
    = 4! \prod_{s=1}^4C_{i_s  i_5}(t_{\sigma(s)} - t_{\sigma(5)}),
\end{multline}
and Eq.~\eqref{eq:F_sigma_N} can be cast as
\begin{multline}
    F_\sigma = \frac{\beta}{m}  \frac{4!}{3} (a^\dagger a)^4 \\ \sum_{j=1}^N \left[ \prod_{s=1}^4 \left(D_j(t_{\sigma(s)}, t_{\sigma(5)})\right) - (t_{\sigma(4)} \leftrightarrow t_{\sigma(5)}) \right],
\end{multline}
where $D_j(t,t') = \sum_{i=1}^N g_i(t) C_{ij}(t-t')$. Hence the fifth Magnus term is found to read
\begin{multline}\label{eq:GeneralFormula}
    \Theta_5(t) = \left(\frac{-i}{\hbar}\right)^5 \frac{4!}{3} \frac{\beta}{m}  (a^\dagger a)^4 \sum_{\sigma\in S} \lambda_\sigma \\  \int_{(5,t)} dt^5 \sum_{j=1}^N \left[\left(\prod_{s=1}^4 D_j(t_{\sigma(s)},t_{\sigma(5)})\right) - (t_{\sigma(4)} \leftrightarrow t_{\sigma(5)}) \right],
\end{multline}
where one can reduce the problem to the knowledge of the normal modes and frequencies by the explicit formula $D_j(t,t') = i\hbar \sum_{i=1}^N g_i(t) \sum_{k=1}^N O_{ik} O'_{jk} \cos(\omega_k (t-t'))$.

We have thus found an analytical form for the mechanical-state-independent phase factor. As in the single oscillator case, this factor depends on the fourth power of the optical intensity, however the expression now involves the mechanical properties of the body described by its normal mode transformations and associated frequencies. To no surprise, if the potential terms correspond to coupled trapped particles, one recovers Eq.~\eqref{eq:GeneralMagnusFifth} by setting $N=1$, for which the normal mode transformations are trivial. 

Finding how $\Theta_5(t)$ should scale with the number of particles $N$ generally is not evident. In what follows, we will consider once again a pulsed regime where four pulses of light interact with the system, which will consist of many copies of a mechanical oscillator which are coupled to the neighbouring oscillators, and show numerically that there can be superlinar scaling.

\section{Extending the four pulse scheme}\label{sec:Pikovskiextended}
\begin{figure}
    \centering
    \includegraphics[scale=0.67]{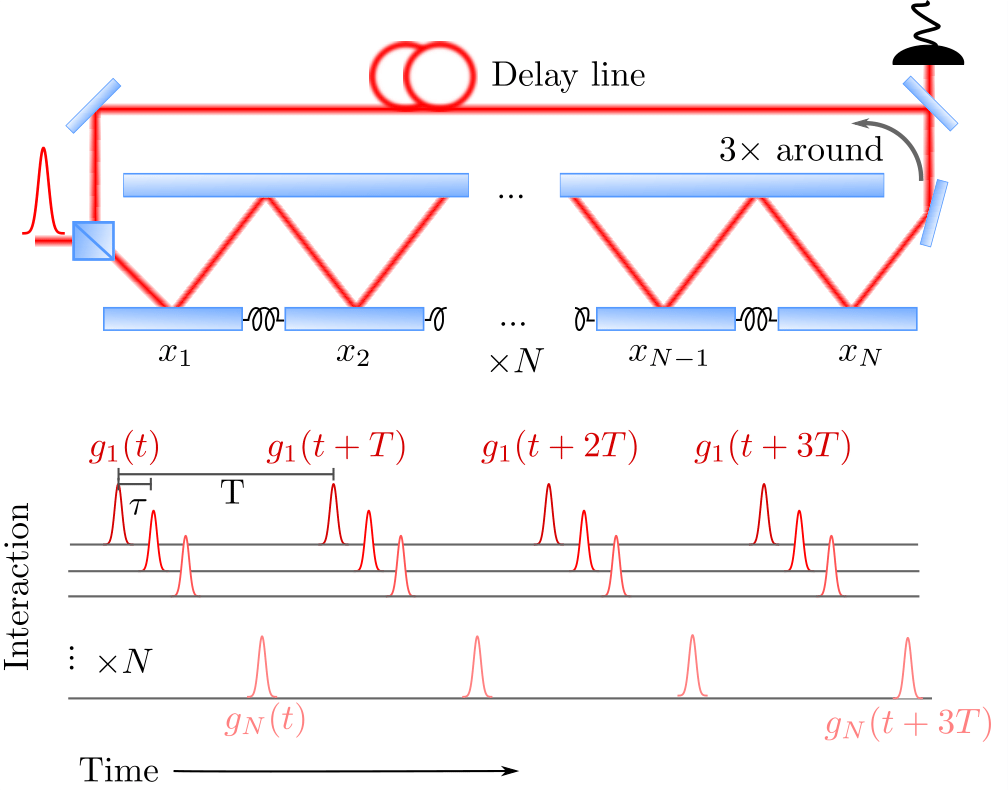}
    \caption{Generalization of the optomechanical protocol where the single trapped particle is replaced with a lattice of coupled oscillators. Each lattice site is affected $4$ times by a light pulse, with time separation $T$. The time taken for the pulse to go from a site to the next is noted $\tau$. The 
    pulse is re-injected only once it has interacted with the whole lattice i.e. $T \geq n\tau$.}
    \label{fig:schematic}
\end{figure}

We seek to use our general formula given by Eq.~\eqref{eq:GeneralFormula} in the case where four pulses of light are sent through a lattice of $N$ coupled oscillators (or sites). That is, we consider coupling functions of the form $g_k(t) = \lambda\sum_{i=0}^3 \delta(t - (t_0 + iT + (k-1)\tau))$ where $\lambda$ is a coupling strength, $T$ is the time separating two pulses for a given site, and $\tau$ the time taken by the pulse to travel from one site to the next one. We shall work under the assumption that $T \geq N\tau$, which means that the pulse is re-injected only once the it has finished interacting with all the lattice sites. The setup and the interaction functions are illustrated in Fig.~\ref{fig:schematic}.

After calculations similar to those carried out in the context of a pulsed regime with a single particle, see Appendix~\ref{sec:extendedproof} for details, one arrives at the following expression for the fifth order Magnus term:
\begin{equation}
    \begin{aligned}
    &\Theta_5(t) = \\
    &\frac{-i}{\hbar} \frac{4!}{3}\frac{\beta}{m}(a^{\dagger}a)^4 \sum_{i_1i_2i_3i_4j=1}^N\sum_{\nu_1\nu_2\nu_3\nu_4=1}^N \left(\prod_{s=1}^4O_{i_s\nu_s}O'_{j\nu_s}\right) \\ &\frac{\lambda^4}{30}\sum_{\alpha_1\alpha_2\alpha_3\alpha_4 = 0}^3\Bigg\{
    H(t, \theta_4,\theta_3,\theta_2,\theta_1)\left(\int_{\theta_3}^{\theta_4}ds - \int_{\theta_4}^t ds\right) \\
 &+ H(t, \theta_1,\theta_4,\theta_3,\theta_2)\left( 4\int_{\theta_3}^{\theta_4}ds - 4\int_{\theta_4}^{\theta_1}ds - \int_{0}^{\theta_3}ds \right) \\
 &+ H(t, \theta_1,\theta_3,\theta_2,\theta_4)\left( \int_{\theta_3}^{\theta_1}ds\right) \\
 &+ H(t, \theta_1,\theta_3,\theta_4,\theta_2)\left( \int_{\theta_3}^{\theta_1}ds\right) \Bigg\}\varphi_{\nu_1\nu_2\nu_3\nu_4}(\theta_1,\theta_2,\theta_3,\theta_4, s),
    \end{aligned}
\end{equation}
where $\varphi_{\nu_1\nu_2\nu_3\nu_4}(t_1,...,t_5) = \prod_{s=1}^4 \cos(\omega_{\nu_s}(t_s-t_5))$ and $\forall r \in \{1,2,3,4\}, \ \theta_r := t_0 + \alpha_rT + (i_r-1)\tau$. 

By fixing $N=1$ one naturally recovers Eq~\eqref{eq:MagnusTermDiracs}. In the case where the coupling between oscillators is zero, then the $O, O'$ normal mode transformation matrices are diagonal, and one is left with the single oscillator phase term multiplied by $N$. Hence, in the uncoupled setting, one expects a linear scaling of the phase contribution coming from the fifth Magnus term with the number of oscillators, as confirmed by numerics.

\begin{figure}
    \centering
    \includegraphics[scale=0.55]{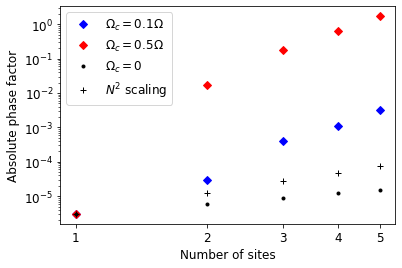}
    \caption{Log-log representation of the absolute phase factor as a function of the number of lattice sites, for different coupling frequencies $\Omega_c$. Trap frequency $\Omega = 2\pi\times 10^5 \text{Hz}$. Delay time $T = \pi/(2\Omega)$ and $\tau = T/(2N)$. The uncoupled case is represented by black dots and scales linearly with the number of lattice sites, while quadratic scaling is represented by black crosses. The absolute phase factor under nonzero coupling is represented with diamonds, and scales superquadratically.}
    \label{fig:superlinear}
\end{figure}

 As shown in Fig.~\ref{fig:superlinear}, numerics reveal the presence of superquadratic scaling, at least up to $N=5$, for coupling frequencies $\Omega_c$ equal to a tenth or a half of the trap frequency $\Omega = 2\pi\times 10^5 \text{Hz}$. The plotted quantity is an absolute phase factor $|\Phi(N)|$ for $N\in\{1,2,3,4,5\}$ defined by $\Theta_5(t) = \frac{-i}{\hbar} \frac{4!}{3}\frac{\beta}{m}(a^{\dagger}a)^4\frac{\lambda^4}{30} \Phi(N)$. In general this is a complicated factor, depending on the coupling $g(t)$, the trapping frequencies of each lattice sites, and the coupling frequencies between them. We have assumed the form given by the last equation, i.e. we work in pulsed regime, and the pulses are separated by a quarter trapping period which is assumed to be the same for all sites, and we have further assumed $\tau = T/(2N)$.
 
\begin{figure}
    \centering
    \includegraphics[scale=0.55]{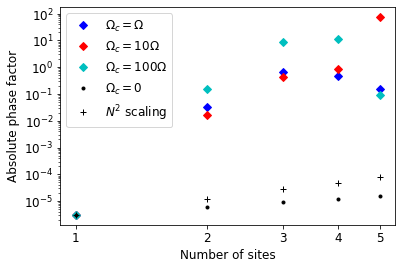}
    \caption{Log-log representation of the absolute phase factor as a function of the number of lattice sites, for different coupling frequencies $\Omega_c$. Trap frequency $\Omega = 2\pi\times 10^5 \text{Hz}$. The uncoupled case is represented by black dots and scales linearly with the number of lattice sites, while quadratic behaviour is represented by black crosses. The absolute phase factor under nonzero coupling is represented with diamonds. There remains an advantage over the uncoupled case, however the scaling with $N$ is not always increasing.}
    \label{fig:bigomegas}
\end{figure}

 For coupling frequencies comparable or higher to the trap frequency, we see in Fig.~\ref{fig:bigomegas} that there remains an advantage, however for the considered coupling frequencies, the absolute phase factor is not an increasing function of the number of sites.

\section{Conclusion}
We have revisited modern approaches to indirectly test non-classical features of spacetime, namely an optomechanical protocol that aims to detect highly nonlinear phase terms that should arise as a result of modifications of commutation relations. Although rigid or uncorrelated macroscopic systems are not viable probes for Planck scale effects, as the GUP corrections to the commutation relations of center of mass variables of such systems are expected to scale as the inverse or squared inverse of the number of constituents, the intermediate case of correlated many-body systems was left unexplored.

In this paper, we have considered such systems in general and, to first order in GUP corrections, we have derived the full dynamics for completely general light-matter coupling functions, and found explicit forms for the fifth order Magnus term which consists of a phase factor that is independent of the mechanical oscillator's state. Many parameters of correlated many-body systems could be tuned, such as the retardation time between two lattice sites, or the spatial width of the optical pulse. Limiting ourselves to a pulsed regime, we have shown that superquadratic scaling is already possible up to $N=5$ subsystems by tuning only the coupling frequency between lattice sites.

This provides new perspectives for controlled many-body quantum dynamics, and may help to re-affirm finely tuned macroscopic probes as potential amplifiers for testing quantum gravitational effects. In addition to showing how macroscopic probes may be relevant to test Planck scale physics, the general calculations laid out in this work may benefit further investigations and tests of fundamental physics using optomechanical systems.

\begin{acknowledgments}
This work was funded by the ESPRC Centre for Doctoral Training in Controlled Quantum Dynamics, and QuantERA ERA-NET Cofund in Quantum Technologies implemented within the European Union's Horizon 2020 Programme and KIAS visiting professorship.
\end{acknowledgments}

\bibliography{ref.bib}

\begin{thebibliography}{22}%
\makeatletter
\providecommand \@ifxundefined [1]{%
 \@ifx{#1\undefined}
}%
\providecommand \@ifnum [1]{%
 \ifnum #1\expandafter \@firstoftwo
 \else \expandafter \@secondoftwo
 \fi
}%
\providecommand \@ifx [1]{%
 \ifx #1\expandafter \@firstoftwo
 \else \expandafter \@secondoftwo
 \fi
}%
\providecommand \natexlab [1]{#1}%
\providecommand \enquote  [1]{``#1''}%
\providecommand \bibnamefont  [1]{#1}%
\providecommand \bibfnamefont [1]{#1}%
\providecommand \citenamefont [1]{#1}%
\providecommand \href@noop [0]{\@secondoftwo}%
\providecommand \href [0]{\begingroup \@sanitize@url \@href}%
\providecommand \@href[1]{\@@startlink{#1}\@@href}%
\providecommand \@@href[1]{\endgroup#1\@@endlink}%
\providecommand \@sanitize@url [0]{\catcode `\\12\catcode `\$12\catcode
  `\&12\catcode `\#12\catcode `\^12\catcode `\_12\catcode `\%12\relax}%
\providecommand \@@startlink[1]{}%
\providecommand \@@endlink[0]{}%
\providecommand \url  [0]{\begingroup\@sanitize@url \@url }%
\providecommand \@url [1]{\endgroup\@href {#1}{\urlprefix }}%
\providecommand \urlprefix  [0]{URL }%
\providecommand \Eprint [0]{\href }%
\providecommand \doibase [0]{http://dx.doi.org/}%
\providecommand \selectlanguage [0]{\@gobble}%
\providecommand \bibinfo  [0]{\@secondoftwo}%
\providecommand \bibfield  [0]{\@secondoftwo}%
\providecommand \translation [1]{[#1]}%
\providecommand \BibitemOpen [0]{}%
\providecommand \bibitemStop [0]{}%
\providecommand \bibitemNoStop [0]{.\EOS\space}%
\providecommand \EOS [0]{\spacefactor3000\relax}%
\providecommand \BibitemShut  [1]{\csname bibitem#1\endcsname}%
\let\auto@bib@innerbib\@empty
\bibitem [{\citenamefont {Garay}(1995)}]{garay1995quantum}%
  \BibitemOpen
  \bibfield  {author} {\bibinfo {author} {\bibfnamefont {L.~J.}\ \bibnamefont
  {Garay}},\ }\href {\doibase 10.1142/S0217751X95000085} {\bibfield  {journal}
  {\bibinfo  {journal} {International Journal of Modern Physics A}\ }\textbf
  {\bibinfo {volume} {10}},\ \bibinfo {pages} {145} (\bibinfo {year}
  {1995})}\BibitemShut {NoStop}%
\bibitem [{\citenamefont {Hossenfelder}(2013)}]{hossenfelder2013minimal}%
  \BibitemOpen
  \bibfield  {author} {\bibinfo {author} {\bibfnamefont {S.}~\bibnamefont
  {Hossenfelder}},\ }\href {\doibase 10.12942/lrr-2013-2} {\bibfield  {journal}
  {\bibinfo  {journal} {Living Reviews in Relativity}\ }\textbf {\bibinfo
  {volume} {16}},\ \bibinfo {pages} {1} (\bibinfo {year} {2013})}\BibitemShut
  {NoStop}%
\bibitem [{\citenamefont {Maggiore}(1993)}]{maggiore1993algebraic}%
  \BibitemOpen
  \bibfield  {author} {\bibinfo {author} {\bibfnamefont {M.}~\bibnamefont
  {Maggiore}},\ }\href {\doibase 10.1016/0370-2693(93)90785-G} {\bibfield
  {journal} {\bibinfo  {journal} {Physics Letters B}\ }\textbf {\bibinfo
  {volume} {319}},\ \bibinfo {pages} {83} (\bibinfo {year} {1993})}\BibitemShut
  {NoStop}%
\bibitem [{\citenamefont {Ali}\ \emph {et~al.}(2009)\citenamefont {Ali},
  \citenamefont {Das},\ and\ \citenamefont {Vagenas}}]{ali2009discreteness}%
  \BibitemOpen
  \bibfield  {author} {\bibinfo {author} {\bibfnamefont {A.~F.}\ \bibnamefont
  {Ali}}, \bibinfo {author} {\bibfnamefont {S.}~\bibnamefont {Das}}, \ and\
  \bibinfo {author} {\bibfnamefont {E.~C.}\ \bibnamefont {Vagenas}},\ }\href
  {\doibase 10.1016/j.physletb.2009.06.061} {\bibfield  {journal} {\bibinfo
  {journal} {Physics Letters B}\ }\textbf {\bibinfo {volume} {678}},\ \bibinfo
  {pages} {497} (\bibinfo {year} {2009})}\BibitemShut {NoStop}%
\bibitem [{\citenamefont {Robertson}(1929)}]{robertson1929uncertainty}%
  \BibitemOpen
  \bibfield  {author} {\bibinfo {author} {\bibfnamefont {H.~P.}\ \bibnamefont
  {Robertson}},\ }\href {\doibase 10.1103/PhysRev.34.163} {\bibfield  {journal}
  {\bibinfo  {journal} {Physical Review}\ }\textbf {\bibinfo {volume} {34}},\
  \bibinfo {pages} {163} (\bibinfo {year} {1929})}\BibitemShut {NoStop}%
\bibitem [{\citenamefont {Kempf}\ \emph {et~al.}(1995)\citenamefont {Kempf},
  \citenamefont {Mangano},\ and\ \citenamefont {Mann}}]{kempf1995hilbert}%
  \BibitemOpen
  \bibfield  {author} {\bibinfo {author} {\bibfnamefont {A.}~\bibnamefont
  {Kempf}}, \bibinfo {author} {\bibfnamefont {G.}~\bibnamefont {Mangano}}, \
  and\ \bibinfo {author} {\bibfnamefont {R.~B.}\ \bibnamefont {Mann}},\ }\href
  {\doibase 10.1103/PhysRevD.52.1108} {\bibfield  {journal} {\bibinfo
  {journal} {Physical Review D}\ }\textbf {\bibinfo {volume} {52}},\ \bibinfo
  {pages} {1108} (\bibinfo {year} {1995})}\BibitemShut {NoStop}%
\bibitem [{\citenamefont {Bawaj}\ \emph {et~al.}(2015)\citenamefont {Bawaj},
  \citenamefont {Biancofiore}, \citenamefont {Bonaldi}, \citenamefont
  {Bonfigli}, \citenamefont {Borrielli}, \citenamefont {Di~Giuseppe},
  \citenamefont {Marconi}, \citenamefont {Marino}, \citenamefont {Natali},
  \citenamefont {Pontin} \emph {et~al.}}]{bawaj2015probing}%
  \BibitemOpen
  \bibfield  {author} {\bibinfo {author} {\bibfnamefont {M.}~\bibnamefont
  {Bawaj}}, \bibinfo {author} {\bibfnamefont {C.}~\bibnamefont {Biancofiore}},
  \bibinfo {author} {\bibfnamefont {M.}~\bibnamefont {Bonaldi}}, \bibinfo
  {author} {\bibfnamefont {F.}~\bibnamefont {Bonfigli}}, \bibinfo {author}
  {\bibfnamefont {A.}~\bibnamefont {Borrielli}}, \bibinfo {author}
  {\bibfnamefont {G.}~\bibnamefont {Di~Giuseppe}}, \bibinfo {author}
  {\bibfnamefont {L.}~\bibnamefont {Marconi}}, \bibinfo {author} {\bibfnamefont
  {F.}~\bibnamefont {Marino}}, \bibinfo {author} {\bibfnamefont
  {R.}~\bibnamefont {Natali}}, \bibinfo {author} {\bibfnamefont
  {A.}~\bibnamefont {Pontin}},  \emph {et~al.},\ }\href {\doibase
  10.1038/ncomms8503} {\bibfield  {journal} {\bibinfo  {journal} {Nature
  communications}\ }\textbf {\bibinfo {volume} {6}},\ \bibinfo {pages} {1}
  (\bibinfo {year} {2015})}\BibitemShut {NoStop}%
\bibitem [{\citenamefont {Bushev}\ \emph {et~al.}(2019)\citenamefont {Bushev},
  \citenamefont {Bourhill}, \citenamefont {Goryachev}, \citenamefont
  {Kukharchyk}, \citenamefont {Ivanov}, \citenamefont {Galliou}, \citenamefont
  {Tobar},\ and\ \citenamefont {Danilishin}}]{bushev2019testing}%
  \BibitemOpen
  \bibfield  {author} {\bibinfo {author} {\bibfnamefont {P.}~\bibnamefont
  {Bushev}}, \bibinfo {author} {\bibfnamefont {J.}~\bibnamefont {Bourhill}},
  \bibinfo {author} {\bibfnamefont {M.}~\bibnamefont {Goryachev}}, \bibinfo
  {author} {\bibfnamefont {N.}~\bibnamefont {Kukharchyk}}, \bibinfo {author}
  {\bibfnamefont {E.}~\bibnamefont {Ivanov}}, \bibinfo {author} {\bibfnamefont
  {S.}~\bibnamefont {Galliou}}, \bibinfo {author} {\bibfnamefont
  {M.}~\bibnamefont {Tobar}}, \ and\ \bibinfo {author} {\bibfnamefont
  {S.}~\bibnamefont {Danilishin}},\ }\href {\doibase
  10.1103/PhysRevD.100.066020} {\bibfield  {journal} {\bibinfo  {journal}
  {Physical Review D}\ }\textbf {\bibinfo {volume} {100}},\ \bibinfo {pages}
  {066020} (\bibinfo {year} {2019})}\BibitemShut {NoStop}%
\bibitem [{\citenamefont {Das}\ and\ \citenamefont
  {Vagenas}(2008)}]{das2008universality}%
  \BibitemOpen
  \bibfield  {author} {\bibinfo {author} {\bibfnamefont {S.}~\bibnamefont
  {Das}}\ and\ \bibinfo {author} {\bibfnamefont {E.~C.}\ \bibnamefont
  {Vagenas}},\ }\href {\doibase 10.1103/PhysRevLett.101.221301} {\bibfield
  {journal} {\bibinfo  {journal} {Physical review letters}\ }\textbf {\bibinfo
  {volume} {101}},\ \bibinfo {pages} {221301} (\bibinfo {year}
  {2008})}\BibitemShut {NoStop}%
\bibitem [{\citenamefont {Penrose}(1996)}]{penrose1996gravity}%
  \BibitemOpen
  \bibfield  {author} {\bibinfo {author} {\bibfnamefont {R.}~\bibnamefont
  {Penrose}},\ }\href {\doibase 10.1007/BF02105068} {\bibfield  {journal}
  {\bibinfo  {journal} {General relativity and gravitation}\ }\textbf {\bibinfo
  {volume} {28}},\ \bibinfo {pages} {581} (\bibinfo {year} {1996})}\BibitemShut
  {NoStop}%
\bibitem [{\citenamefont {Penrose}(2014)}]{penrose2014gravitization}%
  \BibitemOpen
  \bibfield  {author} {\bibinfo {author} {\bibfnamefont {R.}~\bibnamefont
  {Penrose}},\ }\href {\doibase 10.1007/s10701-013-9770-0} {\bibfield
  {journal} {\bibinfo  {journal} {Foundations of Physics}\ }\textbf {\bibinfo
  {volume} {44}},\ \bibinfo {pages} {557} (\bibinfo {year} {2014})}\BibitemShut
  {NoStop}%
\bibitem [{\citenamefont {Diosi}(1987)}]{diosi1987universal}%
  \BibitemOpen
  \bibfield  {author} {\bibinfo {author} {\bibfnamefont {L.}~\bibnamefont
  {Diosi}},\ }\href {\doibase 10.1016/0375-9601(87)90681-5} {\bibfield
  {journal} {\bibinfo  {journal} {Physics letters A}\ }\textbf {\bibinfo
  {volume} {120}},\ \bibinfo {pages} {377} (\bibinfo {year}
  {1987})}\BibitemShut {NoStop}%
\bibitem [{\citenamefont {Bose}\ \emph {et~al.}(2017)\citenamefont {Bose},
  \citenamefont {Mazumdar}, \citenamefont {Morley}, \citenamefont {Ulbricht},
  \citenamefont {Toro{\v{s}}}, \citenamefont {Paternostro}, \citenamefont
  {Geraci}, \citenamefont {Barker}, \citenamefont {Kim},\ and\ \citenamefont
  {Milburn}}]{bose2017spin}%
  \BibitemOpen
  \bibfield  {author} {\bibinfo {author} {\bibfnamefont {S.}~\bibnamefont
  {Bose}}, \bibinfo {author} {\bibfnamefont {A.}~\bibnamefont {Mazumdar}},
  \bibinfo {author} {\bibfnamefont {G.~W.}\ \bibnamefont {Morley}}, \bibinfo
  {author} {\bibfnamefont {H.}~\bibnamefont {Ulbricht}}, \bibinfo {author}
  {\bibfnamefont {M.}~\bibnamefont {Toro{\v{s}}}}, \bibinfo {author}
  {\bibfnamefont {M.}~\bibnamefont {Paternostro}}, \bibinfo {author}
  {\bibfnamefont {A.~A.}\ \bibnamefont {Geraci}}, \bibinfo {author}
  {\bibfnamefont {P.~F.}\ \bibnamefont {Barker}}, \bibinfo {author}
  {\bibfnamefont {M.}~\bibnamefont {Kim}}, \ and\ \bibinfo {author}
  {\bibfnamefont {G.}~\bibnamefont {Milburn}},\ }\href {\doibase
  10.1103/PhysRevLett.119.240401} {\bibfield  {journal} {\bibinfo  {journal}
  {Physical review letters}\ }\textbf {\bibinfo {volume} {119}},\ \bibinfo
  {pages} {240401} (\bibinfo {year} {2017})}\BibitemShut {NoStop}%
\bibitem [{\citenamefont {Marletto}\ and\ \citenamefont
  {Vedral}(2017)}]{marletto2017gravitationally}%
  \BibitemOpen
  \bibfield  {author} {\bibinfo {author} {\bibfnamefont {C.}~\bibnamefont
  {Marletto}}\ and\ \bibinfo {author} {\bibfnamefont {V.}~\bibnamefont
  {Vedral}},\ }\href {\doibase 10.1103/PhysRevLett.119.240402} {\bibfield
  {journal} {\bibinfo  {journal} {Physical review letters}\ }\textbf {\bibinfo
  {volume} {119}},\ \bibinfo {pages} {240402} (\bibinfo {year}
  {2017})}\BibitemShut {NoStop}%
\bibitem [{\citenamefont {Bekenstein}(2012)}]{bekenstein2012tabletop}%
  \BibitemOpen
  \bibfield  {author} {\bibinfo {author} {\bibfnamefont {J.~D.}\ \bibnamefont
  {Bekenstein}},\ }\href {\doibase 10.1103/PhysRevD.86.124040} {\bibfield
  {journal} {\bibinfo  {journal} {Physical Review D}\ }\textbf {\bibinfo
  {volume} {86}},\ \bibinfo {pages} {124040} (\bibinfo {year}
  {2012})}\BibitemShut {NoStop}%
\bibitem [{\citenamefont {Pikovski}\ \emph {et~al.}(2012)\citenamefont
  {Pikovski}, \citenamefont {Vanner}, \citenamefont {Aspelmeyer}, \citenamefont
  {Kim},\ and\ \citenamefont {Brukner}}]{pikovski2012probing}%
  \BibitemOpen
  \bibfield  {author} {\bibinfo {author} {\bibfnamefont {I.}~\bibnamefont
  {Pikovski}}, \bibinfo {author} {\bibfnamefont {M.~R.}\ \bibnamefont
  {Vanner}}, \bibinfo {author} {\bibfnamefont {M.}~\bibnamefont {Aspelmeyer}},
  \bibinfo {author} {\bibfnamefont {M.}~\bibnamefont {Kim}}, \ and\ \bibinfo
  {author} {\bibfnamefont {{\v{C}}.}~\bibnamefont {Brukner}},\ }\href {\doibase
  10.1038/nphys2262} {\bibfield  {journal} {\bibinfo  {journal} {Nature
  Physics}\ }\textbf {\bibinfo {volume} {8}},\ \bibinfo {pages} {393} (\bibinfo
  {year} {2012})}\BibitemShut {NoStop}%
\bibitem [{\citenamefont {Marin}\ \emph {et~al.}(2013)\citenamefont {Marin},
  \citenamefont {Marino}, \citenamefont {Bonaldi}, \citenamefont {Cerdonio},
  \citenamefont {Conti}, \citenamefont {Falferi}, \citenamefont {Mezzena},
  \citenamefont {Ortolan}, \citenamefont {Prodi}, \citenamefont {Taffarello}
  \emph {et~al.}}]{marin2013gravitational}%
  \BibitemOpen
  \bibfield  {author} {\bibinfo {author} {\bibfnamefont {F.}~\bibnamefont
  {Marin}}, \bibinfo {author} {\bibfnamefont {F.}~\bibnamefont {Marino}},
  \bibinfo {author} {\bibfnamefont {M.}~\bibnamefont {Bonaldi}}, \bibinfo
  {author} {\bibfnamefont {M.}~\bibnamefont {Cerdonio}}, \bibinfo {author}
  {\bibfnamefont {L.}~\bibnamefont {Conti}}, \bibinfo {author} {\bibfnamefont
  {P.}~\bibnamefont {Falferi}}, \bibinfo {author} {\bibfnamefont
  {R.}~\bibnamefont {Mezzena}}, \bibinfo {author} {\bibfnamefont
  {A.}~\bibnamefont {Ortolan}}, \bibinfo {author} {\bibfnamefont {G.~A.}\
  \bibnamefont {Prodi}}, \bibinfo {author} {\bibfnamefont {L.}~\bibnamefont
  {Taffarello}},  \emph {et~al.},\ }\href {\doibase 10.1038/nphys2503}
  {\bibfield  {journal} {\bibinfo  {journal} {Nature Physics}\ }\textbf
  {\bibinfo {volume} {9}},\ \bibinfo {pages} {71} (\bibinfo {year}
  {2013})}\BibitemShut {NoStop}%
\bibitem [{\citenamefont {Amelino-Camelia}(2013)}]{amelino2013challenge}%
  \BibitemOpen
  \bibfield  {author} {\bibinfo {author} {\bibfnamefont {G.}~\bibnamefont
  {Amelino-Camelia}},\ }\href {\doibase 10.1103/PhysRevLett.111.101301}
  {\bibfield  {journal} {\bibinfo  {journal} {Physical review letters}\
  }\textbf {\bibinfo {volume} {111}},\ \bibinfo {pages} {101301} (\bibinfo
  {year} {2013})}\BibitemShut {NoStop}%
\bibitem [{\citenamefont {Kumar}\ and\ \citenamefont
  {Plenio}(2020)}]{kumar2020quantum}%
  \BibitemOpen
  \bibfield  {author} {\bibinfo {author} {\bibfnamefont {S.~P.}\ \bibnamefont
  {Kumar}}\ and\ \bibinfo {author} {\bibfnamefont {M.~B.}\ \bibnamefont
  {Plenio}},\ }\href {\doibase 10.1038/s41467-020-17518-5} {\bibfield
  {journal} {\bibinfo  {journal} {Nature communications}\ }\textbf {\bibinfo
  {volume} {11}},\ \bibinfo {pages} {1} (\bibinfo {year} {2020})}\BibitemShut
  {NoStop}%
\bibitem [{\citenamefont {Magnus}(1954)}]{magnus1954exponential}%
  \BibitemOpen
  \bibfield  {author} {\bibinfo {author} {\bibfnamefont {W.}~\bibnamefont
  {Magnus}},\ }\href {\doibase 10.1002/cpa.3160070404} {\bibfield  {journal}
  {\bibinfo  {journal} {Communications on pure and applied mathematics}\
  }\textbf {\bibinfo {volume} {7}},\ \bibinfo {pages} {649} (\bibinfo {year}
  {1954})}\BibitemShut {NoStop}%
\bibitem [{\citenamefont {Brau}(1999)}]{brau1999minimal}%
  \BibitemOpen
  \bibfield  {author} {\bibinfo {author} {\bibfnamefont {F.}~\bibnamefont
  {Brau}},\ }\href {\doibase 10.1088/0305-4470/32/44/308} {\bibfield  {journal}
  {\bibinfo  {journal} {Journal of Physics A: Mathematical and General}\
  }\textbf {\bibinfo {volume} {32}},\ \bibinfo {pages} {7691} (\bibinfo {year}
  {1999})}\BibitemShut {NoStop}%
\bibitem [{\citenamefont {Prato}\ and\ \citenamefont
  {Lamberti}(1997)}]{prato1997note}%
  \BibitemOpen
  \bibfield  {author} {\bibinfo {author} {\bibfnamefont {D.}~\bibnamefont
  {Prato}}\ and\ \bibinfo {author} {\bibfnamefont {P.}~\bibnamefont
  {Lamberti}},\ }\href {\doibase 10.1063/1.473509} {\bibfield  {journal}
  {\bibinfo  {journal} {The Journal of chemical physics}\ }\textbf {\bibinfo
  {volume} {106}},\ \bibinfo {pages} {4640} (\bibinfo {year}
  {1997})}\BibitemShut {NoStop}%
\end{thebibliography}%

\bibliographystyle{apsrev4-1}

\newpage
\onecolumngrid
\begin{appendix}

\section{Detailed expressions for the nested commutators}\label{sec:nestedcommutators}
Recall that the interaction Hamiltonian in the interaction picture reads
\begin{equation}
    \tilde{H}_1(t) = \frac{1}{\sqrt{m\Omega}}g(t)a^{\dagger}a\tilde{q}(t) + \frac{1}{3}m\Omega^2\beta\tilde{p}^4(t).
\end{equation}
Then to first order in $\beta$ one has
\begin{equation}
    \begin{aligned}
     {}   [\tilde{H}_1(t_1),\tilde{H}_1(t_2)] = \frac{1}{m\Omega}(a^{\dagger}a)^2g(t_1)g(t_2)[\tilde{q}(t_1), \tilde{q}(t_2)] + \frac{\sqrt{m\Omega^3}}{3}\beta(a^{\dagger}a)\Big(g(t_1)[\tilde{q}(t_1),\tilde{p}^4(t_2)] + (t_1\leftrightarrow t_2)\Big)
    \end{aligned}
\end{equation}

One can write this explicitly, using the expressions~\eqref{eq:simplecommutators} as
\begin{equation}\label{eq:CommuteOnce}
    \begin{aligned}
     {}  [\tilde{H}_1(t_1),\tilde{H}_1(t_2)] = i\hbar\frac{1}{m\Omega}(a^{\dagger}a)^2&g(t_1)g(t_2)\sin(\Omega(t_2-t_1)) \\ &+ \frac{4}{3}i\hbar \sqrt{m\Omega^3} \beta(a^{\dagger}a)\left(g(t_1)\tilde{p}^3(t_2)\cos(\Omega(t_2-t_1)) - (t_1\leftrightarrow t_2)\right).
    \end{aligned}
\end{equation}
Since the first term commutes with $\tilde{H}_1$, the next nested commutator to first order in $\beta$ takes the simple form
\begin{equation}\label{eq:CommuteTwice}
\begin{aligned}
{}   [\tilde{H}_1(t_1),[\tilde{H}_1(t_2),\tilde{H}_1(t_3)]] = 4(a^{\dagger}a)^2&(i\hbar)^2{\Omega}\beta g(t_1)\\
   &\times \Big(g(t_2)\tilde{p}^2(t_3)\cos(\Omega(t_3-t_1))\cos(\Omega(t_3-t_2)) - (t_2 \leftrightarrow t_3)\Big).
\end{aligned}
\end{equation}
The following term is
\begin{equation}
    \begin{aligned}
    {}   [\tilde{H}_1(t_1),[\tilde{H}_1(t_2)&,[\tilde{H}_1(t_3), \tilde{H}_1(t_4)]]] = 8(a^{\dagger}a)^3(i\hbar)^3\beta \sqrt{\frac{\Omega}{m}} g(t_1)g(t_2) \\
       &\times \Big(g(t_3)\tilde{p}(t_4)\cos(\Omega(t_4-t_1))\cos(\Omega(t_4-t_2))\cos(\Omega(t_4-t_3)) - (t_3 \leftrightarrow t_4)\Big),
    \end{aligned}
\end{equation}
from which one finds Eq.~\eqref{eq:FourNestedCommutator}.
\section{The general fifth order Magnus term}\label{sec:PratoLamberti}
Using the shorthand $\tilde{H}_i := \tilde{H}_1(t_i)$ then it was shown explicitly in~\cite{prato1997note} that
\begin{equation}
    \begin{aligned}
      \Theta_5(t) = \left(\frac{-i}{\hbar}\right)^5\int_{(5,t)}dt^5 \Bigg( & -\frac{1}{30}[\tilde{H}_5, [\tilde{H}_4, [\tilde{H}_3, [\tilde{H}_2, \tilde{H}_1]]]] + \frac{2}{15}[\tilde{H}_1, [\tilde{H}_5, [\tilde{H}_4, [\tilde{H}_2, \tilde{H}_3]]]] -\frac{1}{30}[\tilde{H}_1, [\tilde{H}_4, [\tilde{H}_3, [\tilde{H}_2, \tilde{H}_5]]]] \\
      &-\frac{1}{30}[\tilde{H}_1, [\tilde{H}_5, [\tilde{H}_3, [\tilde{H}_2, \tilde{H}_4]]]] + \frac{1}{15}[[\tilde{H}_5, \tilde{H}_1], [\tilde{H}_4, [\tilde{H}_2, \tilde{H}_3]]] + \frac{1}{15}[[\tilde{H}_4, \tilde{H}_1], [\tilde{H}_5, [\tilde{H}_2, \tilde{H}_3]]] \\
      &- \frac{1}{60}[[\tilde{H}_2, \tilde{H}_3], [\tilde{H}_5, [\tilde{H}_4, \tilde{H}_1]]] + \frac{1}{15}[[\tilde{H}_3, \tilde{H}_1], [\tilde{H}_5, [\tilde{H}_2, \tilde{H}_4]]] - \frac{1}{60}[[\tilde{H}_2, \tilde{H}_4], [\tilde{H}_5, [\tilde{H}_3, \tilde{H}_1]]] \\
      &- \frac{1}{60}[[\tilde{H}_2, \tilde{H}_5], [\tilde{H}_4, [\tilde{H}_3, \tilde{H}_1]]] - \frac{1}{60}[[\tilde{H}_3, \tilde{H}_4], [\tilde{H}_5, [\tilde{H}_2, \tilde{H}_1]]] - \frac{1}{60}[[\tilde{H}_3, \tilde{H}_4], [\tilde{H}_1, [\tilde{H}_2, \tilde{H}_5]]]\\
      &- \frac{1}{60}[[\tilde{H}_5, \tilde{H}_1], [\tilde{H}_3, [\tilde{H}_2, \tilde{H}_4]]] - \frac{1}{60}[[\tilde{H}_4, \tilde{H}_1], [\tilde{H}_3, [\tilde{H}_2, \tilde{H}_5]]] - \frac{1}{60}[[\tilde{H}_3, \tilde{H}_5], [\tilde{H}_4, [\tilde{H}_2, \tilde{H}_1]]]\\
      &- \frac{1}{60}[[\tilde{H}_3, \tilde{H}_5], [\tilde{H}_1, [\tilde{H}_2, \tilde{H}_4]]]- \frac{1}{60}[[\tilde{H}_4, \tilde{H}_5], [\tilde{H}_1, [\tilde{H}_2, \tilde{H}_3]]] - \frac{1}{60}[[\tilde{H}_2, \tilde{H}_3], [\tilde{H}_1, [\tilde{H}_4, \tilde{H}_5]]] \\
      &- \frac{1}{60}[[\tilde{H}_2, \tilde{H}_4], [\tilde{H}_1, [\tilde{H}_3, \tilde{H}_5]]] - \frac{1}{60}[[\tilde{H}_2, \tilde{H}_1], [\tilde{H}_4, [\tilde{H}_3, \tilde{H}_5]]] - \frac{1}{60}[[\tilde{H}_4, \tilde{H}_5], [\tilde{H}_3, [\tilde{H}_2, \tilde{H}_1]]]\\
      &- \frac{1}{60}[[\tilde{H}_3, \tilde{H}_1], [\tilde{H}_4, [\tilde{H}_2, \tilde{H}_5]]] \Bigg).
    \end{aligned}
\end{equation}
As stated in the main text, with the interaction Hamiltonian given by Eq.~\eqref{eq:InteractionHamiltonian}, only the four first terms are linear in the GUP correction $\beta$.

\section{Detailed derivation of the phase term in the pulsed regime}\label{sec:DetailsPikovski}
We begin with Eq~\eqref{eq:GeneralMagnusFifth}
\begin{equation}
    \Theta_5(t) = -\frac{i}{\hbar}\frac{4!}{3}(a^{\dagger}a)^4\frac{\beta}{m}\sum_{i=1}^4\int_{(5,t)}dt^5 \lambda_i\\ \left(\prod_{j=1}^4 g(t_{\sigma_i(j)})\cos(\Omega(t_{\sigma_i(5)} - t_{\sigma_i(j)})) - (t_{\sigma_i(4)} \leftrightarrow t_{\sigma_i(5)}) \right),
\end{equation}
that we rewrite as
\begin{equation}
    \Theta_5(t) = -\frac{i}{\hbar}8(a^{\dagger}a)^4\frac{\beta}{m}( \lambda_1T_1 + \lambda_2T_2 + \lambda_3T_3 + \lambda_4T_4).
\end{equation}
The $\lambda_k$ are known. We seek to express the $T_k$ terms when the interaction $g$ takes the form $g(t) = \lambda\sum_{i=0}^3 \delta(t - \tau_i)$ with no further specification for $\tau_i$. To this end we introduce $f(t_1,...,t_5) := \prod_{j=1}^4\cos(\Omega(t_5-t_j))$. Then
\begin{equation}
\begin{aligned}
    T_k = \lambda^4 \sum_{i_1i_2i_3i_4=0}^3& \Bigg( \int_{(5,t)}dt^5 \delta(t_{\sigma_k(1)} - \tau_{i_1})\delta(t_{\sigma_k(2)} - \tau_{i_2})\delta(t_{\sigma_k(3)} - \tau_{i_3})\delta(t_{\sigma_k(4)} - \tau_{i_4})f(t_{\sigma_k(1)},t_{\sigma_k(2)},t_{\sigma_k(3)},t_{\sigma_k(4)},t_{\sigma_k(5)}) \\
    - &\int_{(5,t)}dt^5 \delta(t_{\sigma_k(1)} - \tau_{i_1})\delta(t_{\sigma_k(2)} - \tau_{i_2})\delta(t_{\sigma_k(3)} - \tau_{i_3})\delta(t_{\sigma_k(5)} - \tau_{i_4})f(t_{\sigma_k(1)},t_{\sigma_k(2)},t_{\sigma_k(3)},t_{\sigma_k(5)},t_{\sigma_k(4)})\Bigg).
\end{aligned}
\end{equation}
Given that $\sigma_1 = (54321)$ we have
\begin{equation}
\begin{aligned}
    T_1 = \lambda^4 \sum_{i_1i_2i_3i_4=0}^3 \Bigg(& \int_{(5,t)}dt^5 \delta(t_{5} - \tau_{i_1})\delta(t_{4} - \tau_{i_2})\delta(t_{3} - \tau_{i_3})\delta(t_{2} - \tau_{i_4})f(t_5, t_4, t_3, t_2, t_1) \\
    &- \int_{(5,t)}dt^5 \delta(t_{5} - \tau_{i_1})\delta(t_{4} - \tau_{i_2})\delta(t_{3} - \tau_{i_3})\delta(t_{1} - \tau_{i_4})f(t_5, t_4, t_3, t_1, t_2)\Bigg).
\end{aligned}
\end{equation}
Integrating the $\delta$ distributions yields
\begin{equation}
\begin{aligned}
    T_1 = \lambda^4 \sum_{i_1i_2i_3i_4=0}^3 H_5(t, \tau_{i_4},\tau_{i_3},\tau_{i_2},\tau_{i_1})\left(\int_{\tau_{i_4}}^t ds - \int_{\tau_{i_3}}^{\tau_{i_4}} ds\right)f(\tau_{i_1},\tau_{i_2},\tau_{i_3},\tau_{i_4}, s).
\end{aligned}
\end{equation}

Next, we have $\sigma_2 = (15423)$ resulting in
\begin{equation}
    \begin{aligned}
    T_2 = \lambda^4 \sum_{i_1i_2i_3i_4=0}^3 \Bigg(& \int_{(5,t)}dt^5 \delta(t_{1} - \tau_{i_1})\delta(t_{5} - \tau_{i_2})\delta(t_{4} - \tau_{i_3})\delta(t_{2} - \tau_{i_4})f(t_1, t_5, t_4, t_2, t_3) \\
    &- \int_{(5,t)}dt^5 \delta(t_{1} - \tau_{i_1})\delta(t_{5} - \tau_{i_2})\delta(t_{4} - \tau_{i_3})\delta(t_{3} - \tau_{i_4})f(t_1, t_5, t_4, t_3, t_2)\Bigg).
\end{aligned}
\end{equation}
Integrating the $\delta$ distributions gives
\begin{equation}
    \begin{aligned}
    T_2 = \lambda^4 \sum_{i_1i_2i_3i_4=0}^3 H_5(t, \tau_{i_1}, \tau_{i_4},\tau_{i_3},\tau_{i_2})\left(\int_{\tau_{i_3}}^{\tau_{i_4}} ds - \int_{\tau_{i_4}}^{\tau_{i_1}} ds\right)f(\tau_{i_1},\tau_{i_2},\tau_{i_3},\tau_{i_4}, s).
\end{aligned}
\end{equation}

The following permutation is $\sigma_3 = (14325)$, and accordingly

\begin{equation}
    \begin{aligned}
    T_3 = \lambda^4 \sum_{i_1i_2i_3i_4=0}^3 \Bigg(& \int_{(5,t)}dt^5 \delta(t_{1} - \tau_{i_1})\delta(t_{4} - \tau_{i_2})\delta(t_{3} - \tau_{i_3})\delta(t_{2} - \tau_{i_4})f(t_1, t_4, t_3, t_2, t_5) \\
    &- \int_{(5,t)}dt^5 \delta(t_{1} - \tau_{i_1})\delta(t_{4} - \tau_{i_2})\delta(t_{3} - \tau_{i_3})\delta(t_{5} - \tau_{i_4})f(t_1, t_4, t_3, t_5, t_2)\Bigg),
\end{aligned}
\end{equation}
such that one arrives at
\begin{equation}
    \begin{aligned}
    T_3 = \lambda^4 \sum_{i_1i_2i_3i_4=0}^3 \left(H_5(t, \tau_{i_1}, \tau_{i_4},\tau_{i_3},\tau_{i_2})\int_{0}^{\tau_{i_2}} ds - H_5(t, \tau_{i_1}, \tau_{i_3},\tau_{i_2},\tau_{i_4})\int_{\tau_{i_3}}^{\tau_{i_1}} ds\right)f(\tau_{i_1},\tau_{i_2},\tau_{i_3},\tau_{i_4}, s).
\end{aligned}
\end{equation}
Finally, from $\sigma_4 = (15324)$ it follows that

\begin{equation}
    \begin{aligned}
    T_4 = \lambda^4 \sum_{i_1i_2i_3i_4=0}^3 \Bigg(& \int_{(5,t)}dt^5 \delta(t_{1} - \tau_{i_1})\delta(t_{5} - \tau_{i_2})\delta(t_{3} - \tau_{i_3})\delta(t_{2} - \tau_{i_4})f(t_1, t_5, t_3, t_2, t_4) \\
    &- \int_{(5,t)}dt^5 \delta(t_{1} - \tau_{i_1})\delta(t_{5} - \tau_{i_2})\delta(t_{3} - \tau_{i_3})\delta(t_{4} - \tau_{i_4})f(t_1, t_5, t_3, t_4, t_2)\Bigg),
\end{aligned}
\end{equation}
which results in

\begin{equation}
    \begin{aligned}
    T_4 = \lambda^4 \sum_{i_1i_2i_3i_4=0}^3 \left(H_5(t, \tau_{i_1}, \tau_{i_4},\tau_{i_3},\tau_{i_2})\int_{\tau_{i_2}}^{\tau_{i_3}} ds - H_5(t, \tau_{i_1}, \tau_{i_3},\tau_{i_4},\tau_{i_2})\int_{\tau_{i_3}}^{\tau_{i_1}} ds\right)f(\tau_{i_1},\tau_{i_2},\tau_{i_3},\tau_{i_4}, s).
\end{aligned}
\end{equation}
Summing up these expressions, noting that $\lambda_1 = \lambda_3 = \lambda_4 = -\frac{1}{30}$ and $\lambda_2 = \frac{4}{30}$ one arrives at
\begin{equation}\label{eq:G12}
\begin{aligned}
 \sum_{k=1}^4 \lambda_kT_k = \frac{\lambda^4}{30}\sum_{i_1i_2i_3i_4 = 0}^3 \Bigg\{ &H(t, \tau_{i_4},\tau_{i_3},\tau_{i_2},\tau_{i_1})\left(\int_{\tau_{i_3}}^{\tau_{i_4}}ds - \int_{\tau_{i_4}}^t ds\right) \\
 &+ H(t, \tau_{i_1},\tau_{i_4},\tau_{i_3},\tau_{i_2})\left( 4\int_{\tau_{i_3}}^{\tau_{i_4}}ds - 4\int_{\tau_{i_4}}^{\tau_{i_1}}ds - \int_{0}^{\tau_{i_2}}ds - \int_{\tau_{i_2}}^{\tau_{i_3}}ds\right) \\
 &+ H(t, \tau_{i_1},\tau_{i_3},\tau_{i_2},\tau_{i_4})\left( \int_{\tau_{i_3}}^{\tau_{i_1}}ds\right) \\
 &+ H(t, \tau_{i_1},\tau_{i_3},\tau_{i_4},\tau_{i_2})\left( \int_{\tau_{i_3}}^{\tau_{i_1}}ds\right) \Bigg\}f(\tau_{i_1},\tau_{i_2},\tau_{i_3},\tau_{i_4}, s).
\end{aligned}
\end{equation}

Now let $\tau_k = t_0 + k\theta$ where $\theta = \frac{\pi}{2\Omega} > 0$ and assume $t > t_0 + 3\theta$ then we find
\begin{equation}
\begin{aligned}
 \sum_{k=1}^4 \lambda_kT_k = \frac{\lambda^4}{30}\sum_{i_1i_2i_3i_4 = 0}^3 \Bigg\{ &H(i_4,i_3,i_2,i_1)\left(\int_{t_0+i_3\theta}^{t_0+i_4\theta}ds - \int_{t_0+i_4\theta}^t ds\right) \\
 &+ H(i_1, i_4,i_3,i_2)\left( 4\int_{t_0 +i_3\theta}^{t_0+i_4\theta}ds - 4\int_{t_0+i_4\theta}^{t_0 + i_1\theta}ds - \int_{0}^{t_0+i_2\theta}ds - \int_{t_0+i_2\theta}^{t_0+i_3\theta}ds\right) \\
 &+ H(i_1,i_3,i_2,i_4)\left( \int_{t_0+i_3\theta}^{t_0+i_1\theta}ds\right) \\
 &+ H(i_1,i_3,i_4,i_2)\left( \int_{t_0+i_3\theta}^{t_0+i_1\theta}ds\right) \Bigg\}f(t_0+i_1\theta,t_0+i_2\theta,t_0+i_3\theta,t_0+i_4\theta, s).
\end{aligned}
\end{equation}
One notes that $f(t_0+i_1\theta,t_0+i_2\theta,t_0+i_3\theta,t_0+i_4\theta, s) = \prod_{k=1}^4\cos(\Omega(s-t_0) + \frac{i_k\pi}{2})$ and so with the change of variable $u = \Omega(s-t_0)$ one can write all integrals in the form
\begin{equation}
    \int_{t_0 + \alpha\theta}^{t_0 + \beta\theta}f(t_0+i_1\theta,t_0+i_2\theta,t_0+i_3\theta,t_0+i_4\theta, s) ds = \frac{1}{\Omega}\int_{\alpha\frac{\pi}{2}}^{\beta\frac{\pi}{2}} \prod_{k=1}^4\cos(u - \frac{i_k\pi}{2})du.
\end{equation}

\section{Derivation of the $[\tilde x,\tilde p]$ unequal time commutator}\label{sec:scalarcommutator}

We prove our claim about $[\tilde{x}_i(t_1),\tilde{p}_j(t_2)]$ being a $c$-number and derive the explicit form of $D_j(t,t')$. To this end, one can write the free Hamiltonian $H_0$ as 
\begin{equation}H_0 = \hbar\Omega_F\left(a^{\dagger}a+\frac{1}{2}\right) + \frac{1}{2m}p^Tp + \frac{1}{2}mx^Thx,
\end{equation} 
where $h$ is a symmetric matrix characterising the potential term. One can define normal variables $(\phi,\pi) = (\omega^{\frac{1}{2}}P^T\sqrt{m}x, \omega^{-\frac{1}{2}}P^T\frac{1}{\sqrt{m}}p)$ where $\omega^2 = P^ThP = \text{diag}(\omega_i^2)$ is the diagonal form of the $h$ matrix, consisting of the squares of the normal frequencies, and where $P$ is the orthogonal transformation matrix. Let us stress that $m$ is a scalar and $\omega$ is a diagonal matrix. With those normal variables, the free Hamiltonian takes the simple form
\begin{equation}H_0 = \hbar\Omega_F\left(a^{\dagger}a+\frac{1}{2}\right) + \frac{1}{2}\sum_{i=1}^N \omega_i(\pi_i^2 + \phi_i^2).\end{equation}
We furthermore note that the transformation is canonical as $[\phi_i,\pi_j] = i\hbar\sqrt{\omega_i/\omega_j}\sum_{k,l=1}^NP_{ki}P_{lj}\delta_{k,l} = i\hbar\delta_{ij}$ by the fact that $P^TP = I_N$. For simplicity we will rewrite the variable transformations as $x = O\phi$ and $p = O'\pi$ where $O$ and $O'$ are orthogonal matrices. Using the shorthand notation $U_0(t) := e^{-iH_0t/\hbar}$ one can now establish that
\begin{equation}
    \begin{aligned}
    {}    [\tilde{x}_i(t_1),\tilde{p}_j(t_2)] &= [U^{\dagger}_{0}(t_1)x_iU_{0}(t_1), U^{\dagger}_{0}(t_2)p_jU_{0}(t_2)]\\
    &= [U^{\dagger}_{0}(t_1)(O\phi)_iU_{0}(t_1), U^{\dagger}_{0}(t_2)(O'\pi)_jU_{0}(t_2)]\\
    &= \sum_{k,\ell = 1}^N O_{ik}O'_{j\ell}[\tilde{\phi}_k(t_1),\tilde{\pi}_{\ell}(t_2)].
    \end{aligned}
\end{equation}
We have $\tilde{\phi}_k(t_1) = \cos(\omega_kt_1)\phi_k + \sin(\omega_kt_1)\pi_k$ and $\tilde{\pi}_{\ell}(t_2) = \cos(\omega_{\ell}t_2)\pi_{\ell} - \sin(\omega_{\ell}t_2)\phi_{\ell}$, from which it follows that
\begin{equation}
    [\tilde{x}_i(t_1),\tilde{p}_j(t_2)] = i\hbar\sum_{k=1}^NO_{ik}O'_{jk}\cos(\omega_k(t_2-t_1)),
\end{equation}
which proves that $C_{ij}(t_1-t_2)$ is a $c$-number.

\section{Extending the optomechanical protocol scheme}\label{sec:extendedproof}
In Eq.~\eqref{eq:GeneralFormula} one can write
\begin{equation}
    \prod_{s=1}^4 D_j(t_{\sigma_s},t_{\sigma_5}) = \sum_{i_1i_2i_3i_4 = 1}^N g_{i_1}(t_{\sigma_1})...g_{i_4}(t_{\sigma_4})f_{i_1i_2i_3i_4}(t_{\sigma_1},...,t_{\sigma_5}),
\end{equation}
where $f_{i_1i_2i_3i_4j}(t_{\sigma_1},...,t_{\sigma_5}) = \prod_{s=1}^4 C_{i_sj}(t_{\sigma_s}-t_{\sigma_5})$. We have $C_{i_sj}(t_{\sigma_s}-t_{\sigma_5}) = i\hbar\sum_{\nu=1}^N O_{i_s\nu}O'_{j\nu}\cos(\omega_{\nu}(t_{\sigma_s} - t_{\sigma_5}))$. Thus one can write
\begin{equation}
    f_{i_1i_2i_3i_4j}(t_{\sigma_1},...,t_{\sigma_5}) = \hbar^4 \sum_{\nu_1\nu_2\nu_3\nu_4=1}^N \left(\prod_{s=1}^4 O_{i_s\nu_s}O'_{j\nu_s}\right)\varphi_{\nu_1\nu_2\nu_3\nu_4}(t_{\sigma_1},...,t_{\sigma_5}),
\end{equation}
where $\varphi_{\nu_1\nu_2\nu_3\nu_4}(t_{\sigma_1},...,t_{\sigma_5}) = \prod_{s=1}^4 \cos(\omega_{\nu_s}(t_{\sigma_s}-t_{\sigma_5}))$. From this, one can express the fifth Magnus term as
\begin{multline}
    \Theta_5(t) = \left(\frac{-i}{\hbar}\right)^5\hbar^4 4!\frac{\beta}{m}(a^{\dagger}a)^4\sum_{\sigma\in S}\lambda_\sigma \sum_{i_1i_2i_3i_4j=1}^N\sum_{\nu_1\nu_2\nu_3\nu_4=1}^N \left(\prod_{s=1}^4O_{i_s\nu_s}O'_{j\nu_s}\right)\times \\
    \int_{(5,t)}dt^5 g_{i_1}(t_{\sigma_1})...g_{i_4}(t_{\sigma_4})\varphi_{\nu_1\nu_2\nu_3\nu_4}(t_{\sigma_1},...,t_{\sigma_5}) - (t_{\sigma_5} \leftrightarrow t_{\sigma_4}).
\end{multline}
We may write 

\begin{equation}
    g_{i_1}(t_{\sigma_1})...g_{i_4}(t_{\sigma_4}) = \lambda^4\sum_{\alpha_1\alpha_2\alpha_3\alpha_4=0}^3 \prod_{r=1}^4 \delta(t_{\sigma_r} - \theta(\alpha_r,i_r)),
\end{equation}
where $\theta(\alpha_r,i_r) := t_0 + \alpha_rT + (i_r-1)\tau$ for convenience.  We shall lighten the notation to $\theta_r$ bearing in mind that it is a function of $\alpha_r$ and $i_r$. The task is now reduced to expressing the integral
\begin{equation}
    \int_{(5,t)}dt^5 \left(\prod_{r=1}^4 \delta(t_{\sigma_r} - \theta_r)\right)\varphi_{\nu_1\nu_2\nu_3\nu_4}(t_{\sigma_1},...,t_{\sigma_5}) - (t_{\sigma_4} \leftrightarrow t_{\sigma_5}).
\end{equation}
After similar calculations as carried out previously, one arrives as
\begin{equation}
    \begin{aligned}
        \sum_{\sigma\in S}&\lambda_\sigma \int_{(5,t)}dt^5 \left(\prod_{r=1}^4 \delta(t_{\sigma_r} - \theta_r)\right)\varphi_{\nu_1\nu_2\nu_3\nu_4}(t_{\sigma_1},...,t_{\sigma_5}) - (t_{\sigma_4} \leftrightarrow t_{\sigma_5})\\
        &= \Bigg\{\lambda_1 H(t,\theta_4,\theta_3,\theta_2,\theta_1)\int_{D_1}ds + \lambda_2 H(t,\theta_1,\theta_4,\theta_3,\theta_2)\int_{D_2}ds\\
        &+\lambda_3 \left( H(t,\theta_1,\theta_4,\theta_3,\theta_2)\int_{D_3} ds + H(t,\theta_1,\theta_3,\theta_2,\theta_4)\int_{D'_3} ds \right)\\
        &+ \lambda_4\left( H(t,\theta_1,\theta_4,\theta_3,\theta_2)\int_{D_4}ds + H(t,\theta_1,\theta_3,\theta_4,\theta_2)\int_{D'_4}ds\right)\Bigg\}\varphi_{\nu_1\nu_2\nu_3\nu_4}(\theta_1, \theta_2, \theta_3, \theta_4, s)
    \end{aligned}
\end{equation}
where the coefficients associated with the permutations are $(\lambda_1,\lambda_2,\lambda_3, \lambda_4) = (\frac{-1}{30},\frac{4}{30},\frac{-1}{30},\frac{-1}{30})$ and the integration domains are given by unions of oriented intervals
\begin{equation}
    \begin{aligned}
    D_1 &= [\theta_4,t]\cup[\theta_4,\theta_3]\\
    D_2 &= [\theta_3,\theta_4]\cup[\theta_1,\theta_4]\\
    D_3 &= [0,\theta_2]; D'_3 =[\theta_1,\theta_3]\\
    D_4 &= [\theta_2,\theta_3]; D'_4 = [\theta_1,\theta_3].
    \end{aligned}
\end{equation}
Thus one can rewrite the fifth Magnus term as
\begin{equation}
    \begin{aligned}
    \Theta_5(t) = -\frac{i}{\hbar} \frac{4!}{3}\frac{\beta}{m}(a^{\dagger}a)^4 \sum_{i_1i_2i_3i_4j=1}^N&\sum_{\nu_1\nu_2\nu_3\nu_4=1}^N \left(\prod_{s=1}^4O_{i_s\nu_s}O'_{j\nu_s}\right) \frac{\lambda^4}{30}\\ \sum_{\alpha_1\alpha_2\alpha_3\alpha_4 = 0}^3\Bigg\{
    &H(t, \theta_4,\theta_3,\theta_2,\theta_1)\left(\int_{\theta_3}^{\theta_4}ds - \int_{\theta_4}^t ds\right) \\
 &+ H(t, \theta_1,\theta_4,\theta_3,\theta_2)\left( 4\int_{\theta_3}^{\theta_4}ds - 4\int_{\theta_4}^{\theta_1}ds - \int_{0}^{\theta_2}ds - \int_{\theta_2}^{\theta_3}ds\right) \\
 &+ H(t, \theta_1,\theta_3,\theta_2,\theta_4)\left( \int_{\theta_3}^{\theta_1}ds\right) \\
 &+ H(t, \theta_1,\theta_3,\theta_4,\theta_2)\left( \int_{\theta_3}^{\theta_1}ds\right) \Bigg\}\varphi_{\nu_1\nu_2\nu_3\nu_4}(\theta_1,\theta_2,\theta_3,\theta_4, s).
    \end{aligned}
\end{equation}
Note that with no coupling between the oscillators, $O_{i_s\nu_s}O'_{j\nu_s} = \delta_{i_s,\nu_s}\delta_{j,\nu_s}$ so we are left with 
\begin{equation}
    \begin{aligned}
    \Theta_5(t) = -\frac{i}{\hbar} \frac{4!}{3}\frac{\beta}{m}(a^{\dagger}a)^4 \sum_{j=1}^N& \frac{\lambda^4}{30}\\ \sum_{\alpha_1\alpha_2\alpha_3\alpha_4 = 0}^3\Bigg\{
    &H(t, \theta_4,\theta_3,\theta_2,\theta_1)\left(\int_{\theta_3}^{\theta_4}ds - \int_{\theta_4}^t ds\right) \\
 &+ H(t, \theta_1,\theta_4,\theta_3,\theta_2)\left( 4\int_{\theta_3}^{\theta_4}ds - 4\int_{\theta_4}^{\theta_1}ds - \int_{0}^{\theta_2}ds - \int_{\theta_2}^{\theta_3}ds\right) \\
 &+ H(t, \theta_1,\theta_3,\theta_2,\theta_4)\left( \int_{\theta_3}^{\theta_1}ds\right) \\
 &+ H(t, \theta_1,\theta_3,\theta_4,\theta_2)\left( \int_{\theta_3}^{\theta_1}ds\right) \Bigg\}\varphi_{jjjj}(\theta_1,\theta_2,\theta_3,\theta_4, s).
    \end{aligned}
\end{equation}
The vector of normal frequencies is simply $\omega^T = \Omega(1,1...,1)$ and $\theta_k = t_0 + \alpha_kT + (j-1)\tau$ so the Heaviside functions only depend on the $\alpha$ coefficients. Hence one arrives at $N$ times the Magnus term given by equation~\eqref{eq:G12}, as expected.
\end{appendix}
\end{document}